\documentstyle[11pt,a4wide]{article}
\def\rfnce{\par\noindent\hangindent 20pt {}}
\newcommand{\AAA}[2]{{\it Astr.~Astrophys.\/} {\bf #1}, #2}

\newcommand{\APJ}[2]{{\it Astrophys.~J.\/} {\bf #1}, #2}
\newcommand{\APJSS}[2]{{\it Astrophys.~J.-Suppl.~Ser.\/} {\bf #1}, #2}

\newcommand{\AJ}[2]{{\it Astron.~J.\/} {\bf #1}, #2}

\newcommand{\Nat}[2]{{\it Nature\/} {\bf #1}, #2}
\newcommand{\MNRAS}[2]{{\it Mon.~Not.~R.~astr.~Soc.\/} {\bf #1}, #2}

\begin{document}

\title{On the formation of elliptical galaxies \\ [30 mm]}

\author{A. Burkert \\
      Max-Planck-Institut f\"ur Astrophysik \\
      Karl-Schwarzschild-Str.\ 1, 85740 Garching bei M\"unchen \\
      Fed.\ Rep.\ of Germany \\ [10 mm]
      email: andi@ibm-1.mpa-garching.mpg.de \\ [50 mm]}

\date{submitted to {\em Mon. Not. R. Astr. Soc.}}

\maketitle
\newpage
\begin{abstract}

   The violent relaxation of dissipationless stellar systems as a possible
   scenario
   for the formation of
   elliptical galaxies is investigated
   in detail. It is shown that a cold dissipationless collapse of initially
   spherically symmetric or irregular stellar systems
   leads to universal de Vaucouleurs profiles only outside 1.5 effective
   radii $R_e$. Inside $1.5 R_e$ the surface density profiles depend
   strongly on the initial conditions and are in general not in
   agreement with the de Vaucouleurs law. This result is in contradiction
   to
   the observations which show that all elliptical galaxies have
   de Vaucouleurs profiles inside $1.5 R_e$ and show strong deviations
   from such a universal surface brightness distribution outside $1.5 R_e$.
   The $r^{1/4}$-profiles of E-galaxies are therefore not a direct result
   of collisionless violent relaxation and angular momentum redistribution.
   It is rather a universal initial density profile
   which is required in order to explain the universal structure of
   E-galaxies.

   A new model for the formation of E-galaxies is presented which
   assumes that the
   ellipticals and the bulges of spiral galaxies
   form from
   isothermal, gaseous spheres which accumulated in the core regions
   of dark matter halos and condense into stars through a cooling
   instability. It is shown that under these circumstances
   a stellar system would form with a
   de Vaucouleurs profile in the observed radius range, that is
   between $0.1 R_e \leq R \leq 1.5 R_e$. This scenario predicts that
   early type galaxies
   trace directly the core radii and core masses
   of their dark matter halos, providing important information on the
   nature
   of the dark matter component in galaxies.

\end{abstract}

\section{Introduction}

Two different theories for the formation of elliptical galaxies are
currently discussed in the
literature: the cold dissipationless collapse and the merging scenario. The
cold collapse model (van Albada 1982) assumes that ellipticals formed from
an initially gaseous protogalaxy which dissipated
its internal kinetic and thermal energy and began to collapse. A subsequent
violent star burst transformed
most of the gas into stars on a timescale, shorter than the
timescale which was required in order for the gas to settle
into a rotationally supported, disk-like substructure. The merging scenario
(Toomre 1977), on the other hand, assumes that primordial
density fluctuations first formed spiral galaxies.
These spirals lateron  merged, resulting in spheroidal ellipticals as a
secondary event.

The collisionless merging of spiral galaxies has recently been
studied in detail by Hernquist (1992, 1993). Hernquist shows that
the merger remnants have central densities which are too low
when compared with elliptical galaxies. This result leads to the
conclusion that the merging spirals either contained already massive
bulges (Hernquist 1993) or that ellipticals formed from preferentially
gaseous
protodisks (Steinmetz \& M\"uller 1993). In the first case, the problem of
the formation of early type galaxies is not solved as one would still have
to
explain the formation of massive, dense stellar bulges in spiral galaxies
and one has to understand, why only Sa-type galaxies merged and formed
E-galaxies.  In the second, more likely case the merging leads again
to a dissipation and radial contraction phase of the gas inside the merger
remnant.  This scenario is therefore very similar to
the cold collapse model, assuming very special, irregular initial
conditions.

One of the most important constraints for models of E-galaxy formation are
their universal surface brightness profiles which were discovered by de
Vaucouleurs (1948). The universality of these de Vaucouleurs profiles
indicates
that there exists a relaxation mechanism which efficiently erases the
information about the details of the initial state, leading to similar final
states. Note,
that in this case the density structure of E-galaxies cannot be used in
order to
reconstruct the initial conditions which lead to the formation of E-galaxies.

It is generally assumed that the dominant relaxation mechanism in ellipticals
is collisionless violent relaxation (Lynden-Bell 1967) and angular momentum
redistribution.  The main requirement is a gravitational potential which
varies with large amplitudes on timescales similar to the dynamical timescale.
Then, the specific energy of the stars is not conserved and energy
redistribution
could lead to an universal final state which is independent of the initial
stellar
energy distribution. In fact both, merging and a violent collapse lead to
strong
variations of the gravitational potential, required for violent relaxation
to be effective.

Van Albada (1982), in a famous and pioneering paper discussed the violent
relaxation of stellar systems and discovered that the final equilibrium
state depends strongly on the initial virial coefficient
$\eta_{vir} = 2 E_{kin}/ |E_{pot}|$ where $E_{kin}$ is the total kinetic
energy and $E_{pot}$ is the potential energy of the system. In initially
virialized
stellar systems ($\eta_{vir} \approx 1$) the stellar energy distribution
is almost
conserved. Therefore the final equilibrium state depends strongly
on the initial state and is general not in agreement with the observed
structure of E-galaxies. On the other hand, cold stellar systems
($\eta_{vir} \leq 0.1$) form anisotropic, universal equilibrium states
which
have
de Vaucouleurs profiles.  This result seems to confirm the importance of
violent
relaxation for the formation of E-galaxies.

Since van Albada's work several authors have investigated the cold,
dissipationless collapse of stellar systems in great detail (May \& van
Albada 1984; Aguilar \& Merritt 1990; Burkert 1990;
Londrillo et al. 1991; Cannizzo \& Hollister 1992). May and
van Albada (1984) argued that clumpy initial conditions lead
to surface density profiles which fit the $r^{1/4}$-law even better than
initially spherically symmetric models. This would indicate that
ellipticals formed through the violent merging of stellar subunits
with
smaller masses ($M \approx 10^9 M_{\odot}$), in agreement with
cosmological hierarchical merging scenarios. On the other hand,
Aguilar and Merritt (1990) studied spherically symmetric and centrally
condensed initial conditions and found, that cold stellar systems
with an initial density profile $\rho(r) \sim r^{-1}$ achieve
$r^{1/4}$-profiles with a similar accuracy as the clumpy initial
conditions.
In addition, analytical distribution functions have been invented
( Bertin
\& Stiavelli 1984; Bertin, Saglia \& Stiavelli 1988,1992, Hjorth \& Madsen
1991) which fit the brightness profiles of ellipticals and the physics
of violent relaxation and phase mixing has been investigated in detail
(Tremaine, H\'{e}non \& Lynden-Bell 1986;
Carlberg 1986; Spergel \& Hernquist 1992).

In most of these articles, the basic
constraint was a resonably good fit to the de Vaucouleurs profile.
Indeed, the published surface density distributions of violent collapse
models fit the $r^{1/4}$-law quite
well, however only in a limited radius range. Londrillo et al. (1991),
for example, note
that the models of van Albada (1982) and Aguilar and Merritt (1990) show
significant departures from the de Vaucouleurs law in the inner and outer
regions. Note that also the observations of ellipticals show departures
from the $r^{1/4}$-law in certain radius ranges. Therefore the important
question
arises whether the region where ellipticals have universal brightness
profiles matches the region where violent relaxation leads to such a
profile.  It is this question which will be examined in this paper.

Section 2 presents a new method which can be used in order
determine whether a numerically derived surface density profile agrees
with the universal
brightness profiles of elliptical galaxies.
Section 3 shows that, in general, a cold dissipationless collapse and by
this violent relaxation and angular momentum redistribution does not
lead to surface density profiles which
agree with the observations. This result leads to the conclusion that
there must exist another, more effective mechanism which explains
the universality of E-galaxies.
Section 4 presents a new scenario for the formation of
E-galaxies.

\section{Do ellipticals have $r^{1/4}$-law profiles?}

De Vaucouleurs (1948) found that elliptical galaxies have universal surface
density profiles which can be well described by the empirical relation

\begin{equation}
   \log \left( \frac{\Sigma (R)}{\Sigma_e} \right) = -3.331 \left[ \left(
      \frac{R}{R_e} \right)^{1/4} - 1 \right].
\end{equation}

$\Sigma$ is the surface density, assuming a constant mass-to-light ratio.
$R$ is the projected distance from the center of the galaxy. $R_e$
represents the
effective radius which for a perfect $r^{1/4}$-law would determine the
isophote that contains half of the total mass, and $\Sigma_e = \Sigma (R_e)$.

Since de Vaucouleurs' work, the universality of the
brightness profiles of elliptical galaxies has been confirmed by many
additional
observations (see e.g. Schombert 1986, 1987; Caon, Capaccioli \& Rampazzo
1990). It was however also demonstrated that E-galaxies often show
strong deviations from the $r^{1/4}$-law.
As elliptical galaxies, as well as theoretical violent relaxation models, do
not lead to perfect de Vaucouleurs profiles, $R_e$ looses its
meaning as the projected half-mass radius and should be considered
as a free parameter which specifies the scale length of the best fitting
$r^{1/4}$-law
within a given radius range.  Note that there exist observed surface
brightness
profiles of E-galaxies which can be fitted by $r^{1/4}$-profiles with
different corresponding effective radii, depending on the radius range
(Schombert 1986, 1987).

In order to specify in which radius range and with which accuracy
ellipticals do follow a universal profile
Burkert (1993) investigated the photometric surface brightness
profiles of a large sample of elliptical galaxies. He found that all
ellipticals,
independent of their global parameters, like their luminosity,
ellipticity and
anisotropy, fit the de Vaucouleurs law remarkably well
within the same effective radius range $0.1 R_e \leq R \leq 1.5 R_e$.
The scale length $R_e$ must be defined consistently, using the slope of
the best
fitting $r^{1/4}$-profile $\Sigma_{1/4}(R)$ through the observational
data points within the above specified radius range:

\begin{equation}
   R_e = \left( \frac{-3.331}{d \log (\Sigma_{1/4})/dR^{1/4}} \right)^4 .
\end{equation}

The scale length $R_e$ of the best fitting $r^{1/4}$-law, determined
through equation 2, can deviate from the projected half-light radius
within a factor of 2.
Note also, that the radius range within which the data shows a
universal profile
covers more than 60 \% of the total luminous mass and therefore is
representative for the stellar distribution in these galaxies.
All analyzed ellipticals show a maximum
deviation of the observed surface density distribution $\Sigma (R)$
from the perfect $r^{1/4}$-law within $0.1 R_e \leq R \leq 1.5 R_e$
which is smaller than

\begin{equation}
\Delta \log \Sigma = \log \Sigma - \log \Sigma_{1/4} \leq 0.08.
\end{equation}

As the surface density changes within this
region by $\log \Sigma (R=0.1 R_e)-\log \Sigma (R=1.5 R_e)=1.7$, the
deviations are smaller
than $5 \%$ of the total logarithmical surface density change. In addition,
it was shown that
ellipticals, in general, do not have de Vaucouleurs-like surface density
profiles outside this
radius range (see also Schombert 1986, 1987). We therefore can conclude
that only those
theoretical models of E-galaxy formation will reproduce the observed
universal surface density
profiles of E-galaxies which fit the $r^{1/4}$-law in the radius range
$0.1 R_e \leq R \leq 1.5 R_e$ with
a similar high accuracy. In particular, numerical
models which show a the $r^{1/4}$-law only in another effective radius range
should be rejected.

In the following analysis the numerical results of violent relaxation
and angular momentum redistribution will be compared with
these new observations.
First, the correct effective radius range has to be found. This is done
by calculating $\log \Sigma (R^{1/4})$ from the projected particle
distribution.
Then, the radius range $0.6 \leq (R/R_e)^{1/4} \leq 1.1$ is specified
iteratively, with $R_e$ being
defined consistently through the slope $b$ of the best fitting straight line
$log \Sigma_{1/4}(R) = a - b \times R^{1/4}$ through the data within this
radius range, using the relation: $(R_e)^{1/4} = -3.331/b$. Finally, the
maximum
deviation $\Delta log \Sigma$ (equation 3) is determined and compared
with the observations.

\section{The surface density profiles of cold collapse models.}

Does violent relaxation
fulfill the constraint, imposed by equation 3?
In order to answer this question we reinvestigate some of the collapse
calculations which have been frequently discussed in the literature.
In the following, we assume that the stellar system
has a total mass
$M = 1$, an initial radius $R_{max} = 1$ and a gravitational constant
$G = 1$, leading to a mean initial density $\overline{\rho} = 0.24$
and an initial free fall time $\tau_{ff} = 1.11$.
The N-body calculations are performed using
a vectorized version of the Barnes-Hut TREECODE (Barnes \& Hut 1986)
which was written by Lars Hernquist (Hernquist 1987). The code
parameters (Hernquist 1988) are set to: tolerance parameter
$\theta = 0.7$, timestep $\Delta t=0.005 \tau_{ff}$ and force computation
with quadrupole terms.  In order to resolve the interesting radius range
between
$0.1 R_e$ and $1.5 R_e$ the smoothing length $\epsilon$ has to be much
smaller than $0.1 R_e$. Such a small $\epsilon$ requires a large number $N$ of
test particles. Otherwise, near encounters
would affect the dynamics.
Test calculations show that a total number of
$N = 5~10^4 - 10^5$ particles and a smoothing length of $\epsilon = 0.005
\approx 0.02 R_e$ is
good enough in order to achieve the required resolution without affecting
the global dynamics and
conserving the total energy of the system better than one percent.

The calculations are stopped when an equilibrium state is reached
(typically after
$5 - 10 \tau_{ff}$) and the density distribution has become time-independent.
The final equilibrium state is then analyzed using a procedure which
is similar to the observational data reduction algorithm used
by Burkert (1993). First, a line of sight through the equilibrium model
is chosen. Surface density profiles are then determined by binning
the projected particle distribution into centered ellipsoidal
rings of constant surface density.
The inner radius $r_i$ and the outer radius $r_{i+1}$ of the ring
i along the apparent major axis is determined by constructing
 a logarithmical
grid with $r_1 < 0.1 R_e$ and
assuming that $r_{i+1} = 1.1 r_i$. The ellipticity $\epsilon_i$ of
ring $i$ is determined by comparing the surface density distributions
along the apparent major and minor axis. The logarithm of the surface
densities $\log \Sigma_i$ of the rings $i$ is then investigated as a
function of their intermediate axis coordinates $R_i$, defined as

\begin{equation}
   R_i = \left( \frac{0.5(r_i+r_{i+1}) \sqrt{1- \epsilon_i}}{R_e} \right)
\end{equation}

where $R_e$ is the intermediate axis effective radius, defined as
described in section 2.

\subsection {Spherically symmetric collapse models}

We start with an initially homogeneous,
spherically symmetric system. The test particles are distributed
randomly so as to produce a velocity distribution function

\begin{equation}
f(r,v) = \frac{\overline{\rho}}{(2 \pi \sigma^2)^{1.5}} \exp
\left( -0.5 \frac{v^2}{\sigma^2} \right) .
\end{equation}

The velocity dispersion $\sigma$ is determined by the specifying
the initial virial coefficient $\eta_{vir} = 0.1$ (Burkert 1990).

Figure 1 shows the surface density distribution of the final, prolate
equilibrium state.Here the line of
sight is chosen to be parallel to the minor axis of the ellipsoid. In
agreement with the results obtained by
van Albada (1982) and Burkert (1990) the outer region follows
a de Vaucouleurs law whereas the inner part has a characteristic,
polytrope-like structure.
The best fitting $r^{1/4}$-law in the outer region is shown in
Fig. 1 by the straight line.
The model fits the de Vaucouleurs law only outside
$1.5 R_e$, that is at $(R/R_e)^{1/4} > 1.1$ and shows a completely
different profile inside $1.5 R_e$.
This result is in contrast to the observed surface density profiles
of elliptical galaxies, which fit the $r^{1/4}$-law inside $1.5 R_e$
and show strong deviations for $R > 1.5 R_e$. We therefore
have to conclude that this model cannot explain the present structure
of E-galaxies.

The previous result was actually expected from the work done e.g.
by Jaffe (1987), Tremaine (1987), White (1987) and Hjorth \& Madsen (1991) who
have shown that the distribution function of the stars in the
outer regions of violently
collapsing stellar systems should lead to a density profile
$\rho \sim r^{-4}$.
Such a density distribution provides a reasonably good
fit to the de Vaucouleurs profile outside $2 R_e$, in agreement with
the calculations. Furthermore, the chosen initial conditions have
a low maximum phase space density and therefore form
a large degenerate core (Burkert 1990) which dominates the inner region
and might explain the deviation from a universal surface density distribution
inside $1.5 R_e$.

The effect of the central degeneracy can be strongly reduced if one
starts from centrally condensed systems with regions of initially
high phase space densities. This was demonstrated by Aguilar \& Merritt (1990)
who studied the evolution of cold stellar systems with an initial
density gradient: $\rho \sim r^{-1}$. The figures 2a and 2b show the
equilibrium surface density profile of a model
with an initial virial coefficient
$\eta_{vir} = 0.1$ and an $r^{-1}$-density profile. Like in the
homogeneous model the outer region ($R > 1.5 R_e$) shows a
de Vaucouleurs profile. From Fig. 2a it is however evident that this
$r^{1/4}$-profile cannot fit the inner region.
Following the procedure which was discussed
in Section 2 we now try to find the correct effective radius range.
Unfortunately the slope $d \log \Sigma /dR^{1/4}$  becomes more negative
with decreasing $R$ and therefore the corresponding effective radius
(equation 2) decreases faster than $R$, leading to an increase of
$(R/R_e)$ with decreasing $R$. This is demonstrated in Figure 2b which
shows that the inner $r^{1/4}$-fit lies at even larger corresponding
effective radii as before.
Only in the innermost region does $(R/R_e)$ decrease. There, it is however
not possible to find a reasonable fit to the de Vaucouleurs profile.

An additional, spherically symmetric cold collapse
calculation with an initial density distribution $\rho \sim r^{-3}$
and $\eta_{vir}=0.1$
shows the same large deviations inside $R_e$ as the previous models.
In total, even for centrally condensed, spherically symmetric
initial conditions
does violent relaxation not lead to the observed universal
surface density profiles of E-galaxies, independent of the initial state.

\subsection{Clumpy initial conditions}

Van Albada (1982) and May and van Albada (1984) argued that the
information about the initial state is erased even more efficiently
in initially
clumpy stellar systems, due to the additional effect of angular
momentum redistribution, leading to a better $r^{1/4}$-profile.

The figures 3a,b,c show the final surface density profile of an initially
clumpy, cold stellar system. Here we adopt the same initial conditions
as in model C3 of van Albada (1982): 20 spheres of radius 0.4 are placed
randomly into a volume of radius 1. The isotropic velocity distribution
of the subunits is chosen such that the virial coefficient is
$\eta_{vir} = 0.1$. The clump masses , positions and velocities are randomly
distributed and each clump is filled homogeneously with stars
which move collectively with the assigned clump velocity.

In this case no dense, polytropic core forms anymore. Again one
can fit a de Vaucouleurs profile to the outer region with $R > 1.5 R_e$
(Fig. 3a).
However, in contrast to the spherically symmetric models,
$d \log \Sigma/dR^{1/4}$ becomes less negative for smaller $R$ and
therefore $(R/R_e)$ increases inwards.
As Fig. 4b shows, the inner region can be fitted nicely by a de
Vaucouleurs law, however now inside $0.1 R_e$ and not between
$0.1 R_e \leq R \leq 1.5 R_e$.
Figure 4c shows the best fit to this intermediate effective radius
range. The maximum deviation from the perfect
$r^{1/4}$-law is $\Delta \log \Sigma = 0.24$ and by this a factor
of 3 larger than the observed maximum deviations
in E-galaxies (equation 3). Note also, that more than $75 \%$ of all
ellipticals in Burkert's sample have even smaller maximum deviations
$\Delta \log \Sigma \leq 0.04$.

Even with this large deviation one might argue, that the violent
relaxation and angular momentum redistribution in clumpy, cold
collapse models
explains the origin of $r^{1/4}$-profiles as also some ellipticals show
systematic deviations, similar to the N-body calculation.
In order to investigate this second order effect, Fig. 3d shows
the normalized slope change

\begin{equation}
  \delta b = \frac{1}{3.331} \left( \frac{d \log \Sigma_1}{d (R/R_e)^{1/4}} -
  \frac{d \log \Sigma_2}{d (R/R_e)^{1/4}} \right)
\end{equation}

as a function of the anisotropy parameter

\begin{equation}
  \left( \frac{v}{\sigma} \right)^* = \frac{v_{rot}/ \sigma}
  {\sqrt{\epsilon /(1- \epsilon)}}
\end{equation}

where $v_{rot}$, $\sigma$ and $\epsilon$ is the maximum rotational
velocity, the
central velocity dispersion and the mean ellipticity, respectively.
In equation (6), $R_e$ is calculated, using equation (2), for the
best fitting $r^{1/4}$-profile within $0.6 \leq (R/R_e)^{1/4} \leq 1.1$.
$\Sigma_1$ and
$\Sigma_2$ denote the best $r^{1/4}$-fits to the data points
within the radius range $0.6 \leq (R/R_e)^{1/4} \leq 0.85$ and
$0.85 \leq (R/R_e)^{1/4} \leq 1.1$,
respectively (for a more detailed discussion, see Burkert 1993).
The parameter $\delta b$ therefore provides an estimate for the
systematic second derivation of $\log \Sigma (R^{1/4})$.
In Fig. 3d the observational
data set, taken from Burkert (1993), is compared with the result of the
clumpy cold collapse model. Anisotropic ellipticals with
$(v/ \sigma)^* < -0.2$ have negative $\delta b$, that is their outer
surface density profiles decrease less steeply with respect to the
$r^{1/4}$-fit of the inner regions.
The theoretical model is anisotropic but has a very large, positive
$\delta b$ and therefore lies in a region which is not populated by
E-galaxies.
Thus, also this clumpy, cold collapse model must be rejected.

Additional calculations of initially cold and clumpy stellar systems
have been performed which all show the same universal structure.
The outer region $R > 1.5 R_e$ of all the models fits nicely
the $r^{1/4}$-law. However inside $1.5 R_e$ the surface
density distribution depends strongly on the chosen initial conditions
and is in general not in a agreement with a de Vaucouleurs law. This
result is in contrast to the observations.
Therefore,
violent relaxation and angular momentum redistribution
alone cannot explain the universal structure of elliptical galaxies.

\subsection{Comparison with former calculations}

It has frequently been claimed in the literature that the cold
collapse models fit
the de Vaucouleurs law very well. Therefore the question arises
whether the present N-body calculations are in conflict with
former computations or whether this disagreement is due to
a different interpretation of otherwise similar results.

Aguilar and Merritt (1990) discussed in great detail the collapse
of spherically symmetric,
cold and centrally condensed initial states. They found that the
$r^{1/4}$-fit becomes better, the colder the initial state, that is the
smaller the initial virial coefficient. Their Fig. 11 however shows clearly
that the de Vaucouleurs profile fits best within $1 R_e \leq R \leq 10 R_e$.
In agreement with the present analysis no fit within
$0.1 R_e \leq R \leq 1.5 R_e$ is found.

A comparison of our clumpy model with van Albada's computation is more
difficult as the initial masses, the positions and the velocities
of the clumps are determined randomly. In fact, the published radial density
distribution of van Albada's model C3 leads to a spherically averaged
surface density profile with a relative slope change
of $\delta b = 0.45$ which is approximately a factor of 2 smaller than
the result, reported in this paper. Still, van Albada's model is located
in the upper left region of figure 3d which is not populated by elliptical
galaxies.
The difference between van Albada's calculation and the present computation
might result from the fact that $\delta b$ depends
strongly on the details of the initial condition, which is
determined randomly.
Test calculations show that different random distributions of the clump
masses, positions, and velocities
lead to variations in $\delta b$ which are
consistent with the difference found between van Albada's model and the
present calculation.

It is important to note that there exist initial conditions which
lead to $r^{1/4}$-profiles within the right radius range. One example
will be discussed in the section 4. Other examples are discussed by
Londrillo et al. (1991) who studied generalized Plummer distributions.
Fig. 4 shows the final surface density profile of a test calculation
with the same initial condition as in model P5 of Londrillo et al..
In perfect agreement
with Londrillo et al., a nice $r^{1/4}$-fit is found within the
radius range $0.6 \leq (R/R_e)^{1/4} \leq 1.1$. This test demonstrates
that the numerical method which is used in this paper
can indeed resolve a de Vaucouleurs profile within the required
radius range,
if such a profile actually forms. On the other hand, the fact that
very special
initial density distributions are required in order to fit
the observed $r^{1/4}$-profiles leads to the conclusion
that the dominant mechanism is not the
statistical mechanics of violent relaxation and angular
momentum redistribution.
The ellipticals instead formed already with the right
density distribution and one has to investigate, which mechanism could
lead to such universal initial conditions.

\section{The central condensation model}

It is well known, that King models (King 1966) fit the
brightness profiles of E-galaxies well, if their concentration indices
$c$ are large: $c = \log (R_t/R_c) \approx 2$.
Here $R_c$ and $R_t$ is the core and tidal radius, respectively.
Early type galaxies might therefore have formed from centrally
concentrated, isothermal
clouds with a cut-off radius $R_c \ll R_t$.
As the stellar component adopts the kinematics and density structure
of the gas inside which it forms, the initial
stellar distribution function is

\begin{equation}
 f(r,v) = \left\{ \begin{array}{ll}
 \frac{\rho_1}{(2 \pi \sigma^2)^{1.5}} \exp \left( \frac{\Phi (r) - 0.5 v^2}
 {\sigma^2} \right) & \mbox{if $r \leq R_t$} \\
 0 & \mbox{otherwise}
 \end{array}
 \right.
\end{equation}

The velocity dispersion $\sigma$ is correlated with the temperature of the
gaseous protogalaxy. The cut-off radius $R_t$ is fixed through a dark
matter halo, as will be discussed lateron.
The figures 5a and 5b
show the equilibrium surface density profiles of two stellar systems
with initial distribution functions, described by equation (8) and
large tidal radii $R_t = 15 R_c$. The model, shown in Fig. 5a, was
initially virialized.
It remained isotropic with some relaxation near $R_t$. The model, shown in
Fig. 5b, had a small initial
velocity dispersion $\sigma$, corresponding to an initial virial coefficient
$\eta_{vir} = 0.1$. This stellar system went through a violent collapse
phase and settled into an anisotropic, prolate equilibrium state.
Both surface density profiles fit the $r^{1/4}$-law well inside the radius
range $0.6 \leq (R/R_e)^{1/4} \leq 1.1$.
They also lie in the $\delta b - (v/\sigma)^*$-region which is populated
by E-galaxies (Fig. 3d).

Up to now, dark matter has been neglected. It is however reasonable to assume,
that the total mass fraction of baryonic to dark matter is independent of
the Hubble type and that, like in spiral galaxies, a dark matter halo dominates
the potential in the outer regions of ellipticals.
Within the framework of the central condensation model we will
assume, that the ellipticals and the bulges of spiral galaxies formed
through the merging of primordial, gaseous density
fluctuations, embedded in dark matter halos.
Whereas the dark matter experienced a collisionless,
violent relaxation phase, the collision dominated gas decoupled
from the dark
matter component, dissipated its energy and fell into the inner regions
of the potential well. Lin and Murray (1992) have shown that
self-regulated high-mass
star formation during this early contraction phase could heat efficiently the
interstellar medium, preventing an early violent star burst. The gas
finally settled into the core region of the dark matter halo
where it achieved an isothermal equilibrium state with sound
velocity $c_g$. In this region one can approximate the dark matter
density distribution by
an isothermal sphere with a velocity dispersion $\sigma _{DM}$.
The gas density distribution $\rho_g$ and therefore the density of
the stellar system
which formed from the gas through a thermal instability
(Field 1965; Murray \& Lin 1989),
is coupled with the distribution of the dark matter $\rho_{DM}$
through the hydrostatic equation:

\begin{equation}
\frac{c_g^2}{\rho_g} \frac{d \rho_g}{dr} = - \bigtriangledown \Phi =
\frac{\sigma_{DM}^2}{\rho_{DM}}
\frac{d \rho_{DM}}{dr}
\end{equation}

where $\Phi$ is the gravitational potential, generated by the
dark matter and the baryons.
The solution to equation (9) is

\begin{equation}
\rho_g = A (\rho_{DM})^{\sigma_{DM}^2/c_g^2}
\end{equation}

where $A$ is a constant.
Given the central gas density $\rho_g(r=0)$ and using equation (10)
one can calculate the radial density distribution of the
baryons and the dark matter, using Poisson's
equation:

\begin{equation}
\frac{1}{r^2} \frac{d}{dr} \left(r^2 \frac{d \ln \rho_g}{dr} \right) =
\frac{4 \pi G}{c_g^2} \left( \rho_g + \rho_{DM} \right) .
\end{equation}

A detailed investigation of the possible solutions for equation (11) as
a function of the free parameters will be presented in a subsequent paper.
As an example, Fig. 6 shows the density distribution of the
baryons and dark matter, as well as the baryonic surface density
distribution for
centrally condensed gaseous spheres with temperature ratios
$\sigma_{DM}^2/c_g^2
\approx 2$. As expected, the baryons show a density distribution which is
equal to the profile of a self-gravitating isothermal sphere in the inner
region
and which has a density cut-off at a radius, equal to the core radius of the
dark matter halo. Such a large, negative density gradient
outside the dark matter core is expected from equation (10) as in the
outer regions $\rho_{DM} \sim r^{-2}$ and therefore $\rho_g \sim r^{-4}$.
Figure
(6) shows that the surface density profile fits the de Vaucouleurs
law nicely within the
radius range $0.1 \leq (R/R_e) \leq 1.5$. Note, that this model also
predicts a strong correlation between the radii and total masses of ellipticals
and the core radii and core masses of their dark matter halos.

\section{Conclusions}

It has been shown that cold, dissipationless collapse models
and therefore violent
relaxation and angular momentum redistribution cannot
explain the universality of the surface brightness profiles of E-galaxies.
These cold collapse models lead to a de Vaucouleurs profile
outside $1.5 R_e$. Inside $1.5 R_e$
the surface density structure depends strongly on the initial
density distribution and does in general not fit an $r^{1/4}$-law.
Ellipticals on the other hand show a de Vaucouleurs law inside $1.5 R_e$ and
no universal surface brightness distribution outside $1.5 R_e$.

As a possible, alternative explanation
we have assumed that the gaseous proto-ellipticals
achieved a universal density profile,
prior to the star formation event.
The central condensation model assumes, that the
protogalactic gas
settled into the core regions of the dark matter halo
and formed an isothermal, hydrostatic sphere, before condensing
into stars through
a cooling instability. Stellar systems which formed in this way
would have the
right surface brightness profiles, observed in elliptical galaxies.
Additional late infall of high angular momentum gas might explain
the formation
of a galactic disk within and around these spheroidal systems which then
would be classified as
bulges of spiral galaxies or as S0-galaxies (Meisels \& Ostriker 1984).
In this scenario, E-galaxies and bulges trace directly the
core radii and core masses of dark matter halos, providing
important information
on the nature of dark matter in galaxies.

\vskip 0.35 in\noindent
{\bf ACKNOWLEDGMENTS}
\vskip 0.18 in\noindent
I thank R. Bender, J. Binney, J. Ostriker and M. Steinmetz for
enlightning discussions.
Special thanks go to L. Hernquist for making available his TREECODE. All
calculations have been performed on the Cray YMP 4/64 of the
Rechenzentrum Garching.

\vfill\eject
\noindent
{\bf REFERENCES}
\medskip
\rfnce{
 Aguilar, L.A. \& Merritt, D., 1990. \APJ{354}{33}.
}\rfnce{
 Barnes, J. \& Hut, P., 1986. \Nat{324}{446}.
}\rfnce{
 Bertin, G. \& Stiavelli, M., 1984. \AAA{137}{26}.
}\rfnce{
 Bertin, G., Saglia, R.B. \& Stiavelli, M., 1988. \APJ{330}{78}.
}\rfnce{
 Bertin, G., Saglia, R.B. \& Stiavelli, M., 1992. \APJ{384}{423}.
}\rfnce{
 Burkert, A., 1990. \MNRAS{247}{152}.
}\rfnce{
 Burkert, A., 1993. {\it Astr.~Astrophys.\/} {\bf 278}, 23.
}\rfnce{
 Cannizzo, J.K. \& Hollister, T.C., 1992. \APJ{400}{58}.
}\rfnce{
 Caon, N., Capaccioli, M. \& Rampazzo, R., 1990. \APJSS{86}{429}.
}\rfnce{
 Carlberg, R.G., 1986. \APJ{310}{593}.
}\rfnce{
 de Vaucouleurs, G., 1948. {\it Ann.~d'Astrophys.\/} {\bf 11}, 28.
}\rfnce{
 Field, G.B., 1965. \APJ{142}{531}.
}\rfnce{
 Hernquist, L., 1987. \APJSS{64}{715}.
}\rfnce{
 Hernquist, L., 1988. {\it Comp. Phys. Comm.} {\bf 48}, 107.
}\rfnce{
 Hernquist, L., 1992. \APJ{400}{460}.
}\rfnce{
 Hernquist, L., 1993. \APJ{409}{548}.
}\rfnce{
 Hjorth, J. \& Madsen, J., 1991. \MNRAS{253}{703}.
}\rfnce{
 Jaffe, W., 1987. In: {\it Structure and Dynamics of Elliptical Galaxies,
 IAU Symp. No. 127}, p.511, ed. de Zeeuw, T., Reidel, Dordrecht.
}\rfnce{
 King, I.R., 1966. \AJ{71}{64}.
}\rfnce{
 Lin, D.N. \& Murray, S.D., 1992. \APJ{394}{523}.
}\rfnce{
 Londrillo, P., Messina, A. \& Stiavelli, M., 1991. \MNRAS{250}{54}.
}\rfnce{
 Lynden-Bell, D., 1967. \MNRAS {136}{101}.
}\rfnce{
 May, A. \& van Albada, T.S., 1984. \MNRAS{209}{15}.
}\rfnce{
 Meisels, A. \& Ostriker, J.P. 1984. \AJ{89}{10}.
}\rfnce{
 Murray, S.D. \& Lin, D.N.C., 1989. \APJ{339}{933}.
}\rfnce{
 Schombert, J.M., 1986. \APJSS{60}{603}.
}\rfnce{
 Schombert, J.M., 1987. \APJSS{64}{643}.
{\bf 45}, L39.
}\rfnce{
 Spergel, D.N. \& Hernquist, L., 1992. \APJ{397}{L75}.
}\rfnce{
 Steinmetz, M \& M\"uller, E., 1993. \AAA{281}{L97}.
}\rfnce{
 Toomre, A., 1977. In: {\it The Evolution of Galaxies and Stellar Populations},
 p.401, ed. Tinsley, B.M. \& Larson, R., New Haven: Yale Univ. Obs..
}\rfnce{
 Tremaine, S., 1987. In: {\it Structure and Dynamics of Elliptical Galaxies,
 IAU Symp. No. 127}, p.367, ed. de Zeeuw, T., Reidel, Dordrecht.
}\rfnce{
 Tremaine, S., H\'{e}non, M. \& Lynden-Bell, D., 1986. \MNRAS{219}{285}.
}\rfnce{
 van Albada, T.S., 1982. \MNRAS{201}{939}.
}\rfnce{
 White, S.D.M., 1987. In: {\it Structure and Dynamics of Elliptical Galaxies,
 IAU Symp. No. 127}, p.339, ed. de Zeeuw, T., Reidel, Dordrecht.
}

\vfill\eject
\centerline{FIGURE CAPTIONS}

\vskip 0.1 in\noindent
{\bf Figure 1.} The logarithm of the equilibrium surface density $\Sigma$,
normalized to the effective surface density $\Sigma_e = \Sigma (R_e)$ is
shown for the initially homogeneous collapse model. $R$ is the intermediate
axis radius and $R_e$ is the effective radius, calculated from the best
$r^{1/4}$-fit to the outer regions (solid line). The error bars show the
statistical $2 \sigma$ errors, calculated from the total number of
test particles within each radial bin.
\vskip 0.07 in\noindent
{\bf Figure 2.} The equilibrium surface density distribution of the initially
centrally concentrated, spherically symmetric collapse model is shown. The
figures 2a and 2b show the best $r^{1/4}$-fit to the outer and inner region,
respectively.
\vskip 0.07in\noindent
{\bf Figure 3.} The figures 3a,b,c show the best $r^{1/4}$-fit to the outer,
inner and intermediate region of the clumpy collapse model. The relative slope
change $\delta b$ of observed ellipticals is shown in Fig. 3d as a function
of their anisotropy parameter $(v/\sigma )^*$. The clumpy model
(starred symbol) lies inside a region which is not populated by galaxies.
The two triangles show the relative slope change of the isotropic and
anisotropic isothermal models, discussed in section 4.
\vskip 0.07in\noindent
{\bf Figure 4.} The equilibrium surface density distribution of a cold
collapse calculation is shown, which started with a polytropic initial
condition
(model P5 of Londrillo et al. (1991)).
\vskip 0.07in\noindent
{\bf Figure 5.} The equilibrium surface density distribution of the isotropic
(Fig. 5a) and anisotropic (Fig. 5b) isothermal models without a dark halo
is compared with the de Vaucouleurs law (solid lines).
\vskip 0.07 in\noindent
{\bf Figure 6.} The density distribution $\rho$ of the baryons (solid line) and
the dark matter halo (dashed line), normalized to the central dark matter
density $\rho_0$ is shown in Fig. 6a as a function of the radius $r$,
normalized
to the dark matter core radius $r_c$. Fig. 6b compares the baryonic surface
density (solid line) with the de Vaucouleurs law (dashed line).

\end{document}